\documentclass{fundam}
\usepackage{url} 
\usepackage[ruled,vlined,linesnumbered]{algorithm2e}
\usepackage{graphicx}
\usepackage{subfigure}

\begin{document}

\setcounter{page}{103}
\publyear{22}
\papernumber{2144}
\volume{188}
\issue{2}

   \finalVersionForARXIV

\title{Adaptive Merging on Phase Change Memory}

\author{Wojciech Macyna\thanks{Address for correspondence: Wroc{\l}aw University of Science and Technology,
               Wybrze\.ze Wyspia\'nskiego 27, 50-370 Wroc{\l}aw, Poland. \newline \newline
                    \vspace*{-6mm}{\scriptsize{Received April 2022; \ accepted December  2022.}}},
              \  Michal Kukowski
\\
 Faculty of Information and Communication Technology \\
 Wroc{\l}aw University of Technology \\
  Wroc{\l}aw, Poland\\
  wojciech.macyna{@}pwr.edu.pl
  michal.kukowski@pwr.edu.pl}

  \maketitle

\runninghead{W. Macyna and M. Kukowski}{Adaptive Merging on Phase Change Memory}

\begin{abstract}
Indexing is a well-known database technique used to facilitate data access and speed up query processing. Nevertheless, the construction and modification of indexes are very expensive. In traditional approaches, all records in the database table are equally covered by the index. It is not effective, since some records may be queried very often and some never. To avoid this problem, adaptive merging has been introduced. The key idea is to create an index adaptively and incrementally as a side-product of query processing. As a result, the database table is indexed partially depending on the query workload.

This paper faces the problem of adaptive merging for phase change memory (PCM). The most important features of this memory type are limited write endurance and high write latency. As a consequence, adaptive merging should be investigated from the scratch. We solve this problem in two steps. First, we apply several PCM optimization techniques to the traditional adaptive merging approach. We prove that the proposed method (eAM) outperforms a traditional approach by 60\%.
After that, we invent the framework for adaptive merging (PAM) and propose a new variant of the PCM-optimized index. It further improves the system performance by 20\% for databases where search queries interleave with data modifications.
\end{abstract}


\section{Introduction}\label{intro}

Database indexing is a very mature technique for accelerating query processing. However, index creation and maintenance can decrease the performance of the database. In traditional approaches, an index covers all records equally. However, in real applications, some records are queried often and some never. Let us assume that the database records selling transactions in the table $Sales$ containing the attributes: $Product$, $Quantity$, $Value$ and $Date$. If the records from the last year are queried, it is not necessary to index the whole table by $Date$. In this case, it would be more profitable to index only the records of the last year and leave the older ones not indexed.

\begin{figure}[ht]
\vspace*{1mm}
\centering
\includegraphics[width=0.6\textwidth]{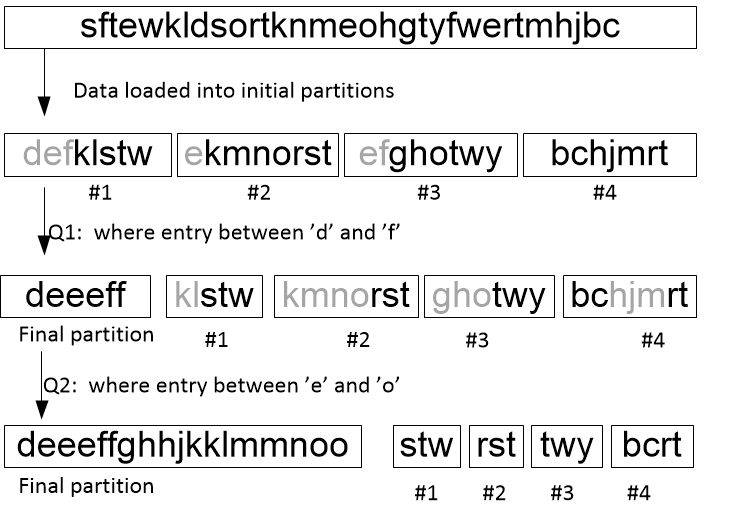}\vspace*{-2mm}
\caption{Adaptive merging}
\label{fig:CRCracking}
\end{figure}

To solve this problem, two main approaches of partially indexing were introduced: database cracking and adaptive merging. In the database cracking (see \cite{Idreos:2007:UCD:1247480.1247527}, \cite{Idreos07databasecracking},\cite{Idreos:2009:STR:1559845.1559878},
\cite{Kersten05crackingthe}), the index is incrementally adapted in response to the actual query workload: only such tables, columns and key ranges inside the table are indexed which are really queried.
In the case of adaptive merging (\cite{Graefe:2010:SSI:1739041.1739087}, \cite{MainMemoryAdaptive}, \cite{IdreosMKG11}, \cite{DBLP:conf/dasfaa/XueQZWY13}) the data are split into $n$ sorted partitions before starting the merging process.
Each merge step affects only such key ranges
that are relevant to actual queries, leaving records in all other key
ranges in their initial places.
Figure \ref{fig:CRCracking} shows an example of adaptive merging.
First, the data entries are loaded into partitions and sorted inside them. Then, the entries fetched from the partitions by query Q1 are stored in the final partition. These entries are depicted using the shadowed font. After that, the query Q2 operates on the partitions updated as a result of Q1.
Subsequent queries may continue indexing until all data entries for the given workload are stored inside the final partition. Thus, indexing is a side effect of the executed queries.

Phase change memory (PCM) is one of the most promising non-volatile storage media. Due to its byte addressability, high access speed and low energy consumption, PCM is the best candidate to cooperate with DRAM as a main memory storage type. However, the drawbacks of PCM are: limited write endurance, high write latency, and asymmetry between read and write operation time. A PCM cell can endure only $10^{7}$ - $10^{8}$ writes. Thus, DRAM cannot be replaced by PCM completely.
Nowadays, the research of PCM is concentrated on many aspects. To prolong the memory lifetime, we need a mechanism that distributes write operations to PCM cells uniformly (see \cite{Qureshi:2009:ELS:1669112.1669117},  \cite{DBLP:conf/waim/OuCXH14}, and \cite{DBLP:conf/ica3pp/WuJY15}). Consequently, many data management techniques must be changed in order to reduce the write traffic (see \cite{RethinkingDatabase},\cite{Viglas:2014:WSJ:2732269.2732277}, and \cite{DBLP:conf/dexa/JurgaM18}).

As PCM is a relatively new storage media, many issues related to databases should be optimized.
The paper faces a problem of adaptive merging on PCM. We resolve the problem in two steps. First, we improve the traditional adaptive merging (AM) by applying several PCM optimization techniques. We call this method: extended adaptive merging (eAM). The detailed descriptions of AM and eAM are presented in subsection \ref{Impl}. Then, we further improve the adaptive merging by inventing a framework that takes into consideration crucial PCM limitations such as: data access asymmetry and limited memory cell endurance. Although the framework can cooperate with any index type, we use a new index called BB+tree. The index is an extended version of DPTree \cite{DBLP:journals/pvldb/ZhouSCHC19} optimized for adaptive merging.
In our model, we consider DRAM as a main memory and PCM as the secondary storage.

\medskip
\textbf{The key contribution} of this paper is the adaptive merging framework described in section \ref{frm}.
Most of the approaches to partial indexing aim at fast and effective index convergence after receiving a set of range queries and do not consider database modification.
Our framework, on the contrary, is very efficient also in the dynamic database environment where updates interleave with range queries.
\textbf{The second contribution} is the adaptation to PCM. We do this by applying some PCM-optimized techniques and proposing a modified variant of PCM index (see section \ref{index}).
Section \ref{exp} presents the experiments which prove the suitability of both the framework and the new techniques for adaptive merging on PCM.

\section{Background}

\subsection{Phase Change Memory (PCM)}

Phase Change Memory (PCM) has attracted attention due to non-volatility, high density, shock-resistivity, and low leakage power. PCM is built out of chalcogenide-based materials. It stores bits by heating a piece of chalcogenide glass with a large current. After cooling down, two forms with different electrical resistance are created: a crystalline state corresponds to "1" and an amorphous state corresponds to "0" \cite{jetcYuXDLC17}.

PCM has been considered as a promising candidate to replace DRAM due to its  lifetime and access latency. Nevertheless, the lifetime of PCM is about $10^9$ times of writes per cell. It is relatively short compared with DRAM. Apart from this, PCM consumes more energy than DRAM for each write. To cope with these problems, the current research is divided into two categories. The first category targets on reducing the total number of writes to PCM. The second category is focused on evenly distributing writes to memory cells. This technique is called wear-leveling.

To reduce the total number of memory writes, a hybrid memory architecture consisting of DRAM and PCM is proposed (see \cite{DBLP:conf/isca/QureshiSR09} and \cite{DBLP:conf/dac/ParkYL11}). Zhou et al. \cite{Zhou:2009:DEE:1555754.1555759} introduce bit-writes removal to avoid unnecessary writes. Namza et al. \cite{DBLP:journals/corr/SohailVV15}  create a DRAM-based system that is used as a cache of PCM. In this way, the write-intensive data could be held in DRAM and the read-intensive one in PCM.

Wear-leveling is a mechanism for extending the lifetime of PCM. Memory blocks should be worn out equally. This is because if one block of PCM wears out, the whole PCM is regarded as broken. Wear-leveling can be implemented at the hardware (see \cite{Zhou:2009:DEE:1555754.1555759}) or operating system level (see \cite{DBLP:conf/rtcsa/PanXHQZ14}). In \cite{DBLP:conf/aspdac/LiuWWSZS13}, Liu et al. provide an application-specific wear-leveling method, which gradually changes the mapping of hot regions in address space to a different part of physical memory. In \cite{DBLP:journals/csur/RashidiJS19}, the most important works about prolonging PCM lifetime are reviewed.

\subsection{Indexing on PCM}

To reduce PCM write traffic, database index structures must be redesigned.

\medskip
In \cite{RethinkingDatabase}, the authors deal with B+tree index optimized for PCM. They present improved algorithms that reduce both execution time and energy on PCM while increasing write endurance.

The authors in \cite{Chi:2014:MBT:2627369.2627630} enhance the performance of B+-tree by introducing unsorted and overflow nodes. The unsorted nodes are used to avoid sorting of items in insert operations. The overflow nodes reduce the write count in the cases where the existing schemes are ineffective.

In \cite{DBLP:conf/adc/LiJYWY16}, the authors are studying indexing on PCM/DRAM-based hybrid memory and propose an improved version of the B+tree called XB+tree (eXtended B+tree). The key idea is to detect the read/write tendency of the index nodes and organize  read-intensive nodes on DRAM while putting write-intensive nodes on PCM. To effectively move nodes between DRAM and PCM, they propose a new node management and migration algorithm. In this way, the read and write operations on PCM are reduced.

The authors in \cite{DBLP:conf/sigmod/OukidLNWL16} propose a hybrid persistent and concurrent Fingering Persistent Tree (FPTree). The FPTree uses fingerprinting, a technique that limits the expected number of in-leaf
probed keys to one. The leaf nodes of FPTree are stored on persistent memory while inner nodes are placed in DRAM.

The paper \cite{DBLP:journals/pvldb/ZhouSCHC19} proposes a new index fully optimized for PCM called DPTree. The index consists of two components: the B+ tree in DRAM and the main index on PCM. First, the operations like insert, delete and update are accumulated in the B+tree. Then, the entries stored in the B+tree are merged into the main index on PCM. Worth mentioning is the fact that the index modification is performed in batches which drastically amortizes persistence overhead. DPTree can work concurrently for multi-core processors and  includes several techniques to achieve crash consistency.

In \cite{DBLP:journals/jise/JabarovP17}, a PCM-variant of the R-tree is proposed. R-tree is a well-known index that can handle spatial data. The proposed PCR-tree has a more compact structure with fewer overlaps.
It reduces the number of PCM writes 30 times in comparison to the traditional R-tree. Moreover, the insert performance is improved by 70\% on average.

The paper \cite{DBLP:conf/dexa/JurgaM18} proposes an implementation of the aggregated R-tree optimized for PCM. The aggregated R-tree extends the original R-tree with the aggregated values connected with the nodes. The proposed method
records the most profitable set of aggregated values in the case when the
memory size is limited. The method chooses to store frequently asked values and
the aggregated values with high calculation cost.

\subsection{Partially indexing} \label{parInd}

In traditional indexing, all data of a particular column or columns are indexed immediately after index creation. Such an approach is not effective in situations when not all data are queried.
Database cracking is an approach where index maintenance is a part of query processing using continuous physical reorganization.
In \cite{Idreos07databasecracking}, the first mature cracking architecture is presented. The approach is implemented as a relational system with minor enhancements to its relational algebra kernel.
In \cite{Idreos:2007:UCD:1247480.1247527}, the authors go beyond the static database. They introduce several novel algorithms for data modification against
a cracked database and prove that the good performance of a cracked database can be maintained in a dynamic environment where updates interleave with queries.
Further improvements are made in \cite{SekPattern} where many different experimental settings, including high and low selectivity queries, and multiple query access patterns are tested.

In \cite{Graefe:2010:SSI:1739041.1739087}, an alternative approach called
adaptive merging is introduced. The main difference is that the data are split to $n$ sorted partitions before starting this process. Each merge step affects only such key ranges that are relevant to actual queries, leaving records in all other key ranges in their initial places.

Both approaches have some weaknesses. Adaptive merging has a high initialization cost but converges rapidly. On the other hand, database cracking enjoys a low initialization cost but converges relatively slowly.  To overcome such limitations, a hybrid approach is proposed (see \cite{IdreosMKG11}). The authors implement several hybrid algorithms for a column-store database system. They  compare a performance of their system against database cracking and adaptive merging, as well as against both traditional full index lookup and scan of unordered data.
Another hybrid approach is proposed in \cite{DBLP:conf/dasfaa/XueQZWY13}. It utilizes a cost model to determine the best operation to refine the index. Experiments show that the hybrid approach can achieve appropriate performance trade-offs between database cracking and adaptive merging.
In \cite{MainMemoryAdaptive} the authors propose three alternative parallel algorithms for adaptive indexing designed for the multi-core system.
The proposition of partial indexing for flash memory can be found in \cite{MacynaK19}. Nevertheless, the authors concentrate on the block addressable memory type which has completely different characteristics than PCM.

Our paper presents a method for adaptive merging for PCM.
The approach is not just limited to optimizing the traditional adaptive merging methods for PCM ( see \cite{IdreosMKG11} and \cite{Graefe:2010:SSI:1739041.1739087}), but it can be treated as a framework where many different index types can be utilized.
We do not take into account database cracking since this approach is not suitable for memory types where the write operation is much more time-consuming than the read one (see \cite{Graefe:2010:SSI:1739041.1739087}).
As a further optimization, we improve DPTree index (\cite{DBLP:journals/pvldb/ZhouSCHC19}) to exploit its potential for bulkloading, which is a dominant operation in adaptive merging.

\section{Adaptive merging framework} \label{frm}

In this section, we propose a new adaptive merging framework (PAM) tailored for phase change memory.

Adaptive merging is a process that creates or refines the database index as a side effect of query execution. Thus, the index is not created for all records of the table at once, but gradually and only for records retrieved by consecutive database queries.
Adaptive merging is applied for the secondary index in databases. So, we consider an $entry$ as a pair: $\langle key, ptr \rangle$, which denote the key value and a pointer to the record, respectively. In the subsequent part of the paper, we will call such an index as "merge index".

\medskip
Two kinds of events may affect the merge index.
\begin{itemize}
\item A search query can invoke adaptive merging process. Intuitively, a query of the key range $<k_1, k_2>$ can add all entries with keys between $k_1$ and $k_2$ to the merge index.
\item As with traditional indexes, operations such as insert, update and delete may affect the merge index. Clearly, adding a record with key $k$ to the database results in inserting an entry into the merge index. On the other hand, removing a record from the database deletes the respective entry from the merge index if the entry has been inserted into it before.
\end{itemize}

\subsection{System concept}

The proposed framework consists of two parts.
The first part is mainly stored on PCM and has the following components:
\begin{itemize}
\item \textit{Partition}. A structure for storing merge index entries. The partitions are created at the beginning of the adaptive merging process. The entries within each partition are sorted by the key. Each partition holds in DRAM the minimal ($min$) and maximal ($max$) key value. So, it is very easy to determine which partitions should be accessed during the particular query execution. When all entries of the partition are moved to the merge index, the partition can be removed from PCM. The adaptive merging process starts with copying all entries to the sorted partitions.
\item \textit{Deletion Journal}. If some data is removed from the database, the merge index entry corresponding to that data must be deleted as well. The \textit{deletion journal} holds the entries deleted from the partition or the entries deleted from the index that was previously copied to it from the partition. The deleted entries must be held on PCM so that they could be restored after the system crash.
\item \textit{Index}. Any index optimized for PCM. Although the framework can work with any index, we propose our version of PDTree that shows very good performance in the adaptive merging process (see section \ref{index}).
\end{itemize}

The second part is stored in DRAM and contains:
\begin{itemize}
\item \textit{Sorting buffer}. The DRAM resident buffer is used to sort the merge index entries before inserting them into the partition. We assume that the maximal partition size should not exceed the sorting buffer size.
\item \textit{Insertion Journal}. It holds a list of key ranges already copied  by the query from the partitions to the merge index. Such values may be stored in DRAM because they can be easily restored after the system crash using the merge index.
To speed up access to the insertion journal, the values are stored in a tree-like structure. For simplicity, we do not reflect this fact in the subsequent description.
\item \textit{Partition set}. The partition set is used to manage the partitions. It contains a list of partitions with their attributes. The attributes depict the minimal ($min$) and maximal ($max$) key values inside the partition.
In this way, it is very easy to identify which partition holds the required data.
Apart from that, the memory
addresses of the first and the last entry in the partition are stored to speed up the memory search. If all entries from the partition are copied to the merge  index, the partition is removed from PCM.
\end{itemize}

\begin{figure}[h]
\centering
\includegraphics[width=0.72\textwidth]{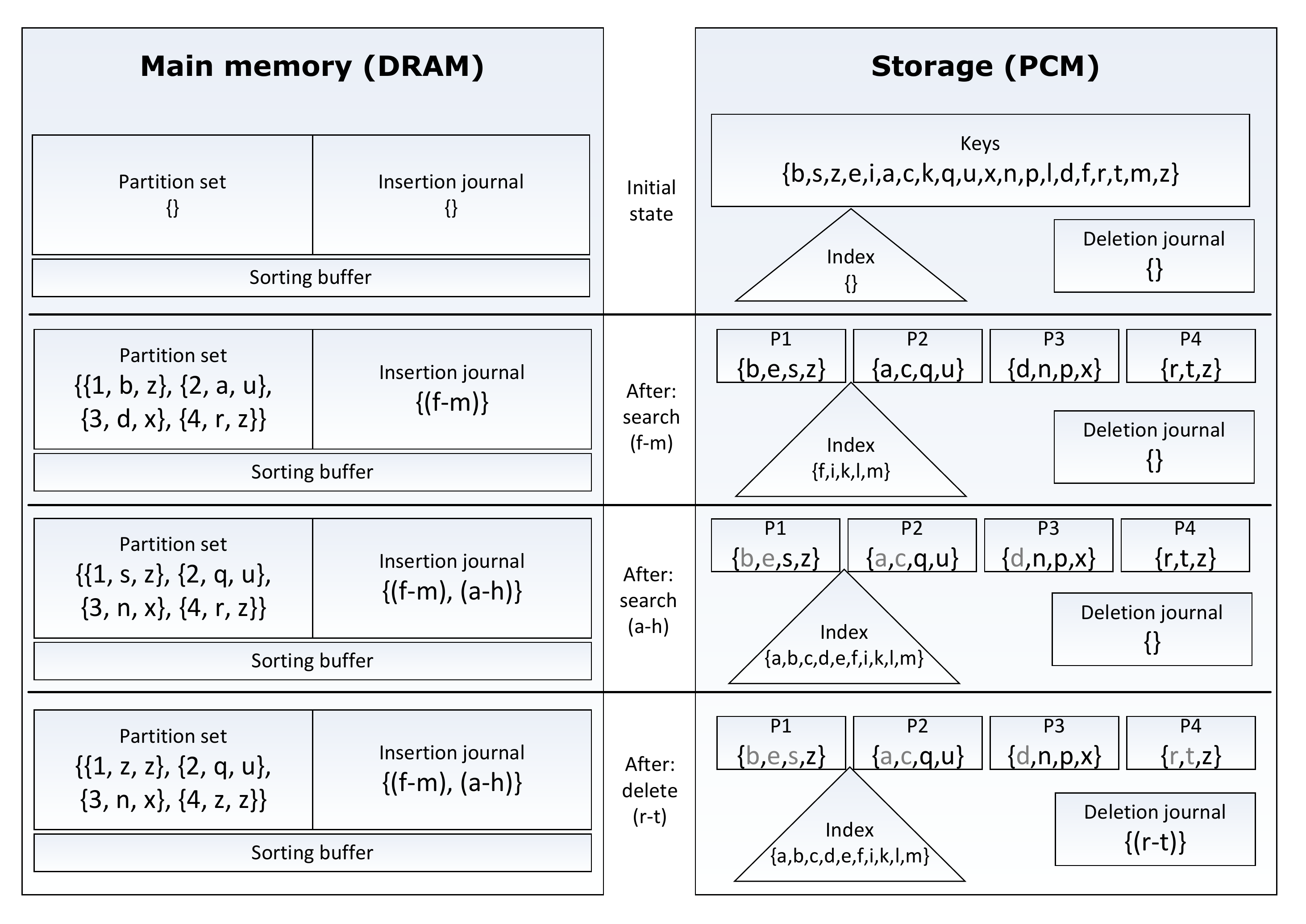}\vspace*{-2mm}
\caption{Adaptive merging for PCM}
\label{fig:figure31}
\end{figure}

Figure \ref{fig:figure31} shows an example of our method. Let us consider a database table containing entries with the following keys: ($b$, $s$, $z$, $e$, $i$, $a$, $c$, $k$, $q$, $u$, $x$, $n$, $p$, $l$, $d$, $f$, $r$, $t$, $m$, $z$).
In the initial state, the structures like the merge index, deletion and insertion journal, and the partition set are empty.
Let us assume that we receive a query of the range $f-m$. The entries are sorted in the sorting buffer and stored in the partitions: $P1$, $P2$, $P3$, $P4$. The selected entries $f$,$i$,$k$,$l$,$m$ are added to the merge index and the range $f-m$ is stored in the insertion journal. Additionally, the partition set is updated: it holds a partition $id$ and the minimal and maximal key value in the partition. As we mentioned above, the partition set holds also the PCM addresses of the first and last entry in the partition. For simplicity, we do not show it in the example.
In the second query, the values between $a$ and $h$ are selected. In that case, the content of the particular partitions changes. The values: $b$ and $e$ from $P1$, $a$ and $c$ from $P2$, and $d$ from $P3$ are copied to the merge index and become invalid. This is marked using the shadowed font. Please note that $P4$ remains intact since it contains data out of the required range. Moreover, the values $f$, $g$, $h$ are inside the insertion journal. It means that they can be fetched directly from the merge index by subsequent queries. After the query execution, a new range $a-h$ is added to the insertion journal.
The last query is an example of a delete operation. The deleted values of the range $r-t$ are removed from $P4$ and added to the deletion journal.

As we can see in this example, the merge index is created as a side-product of query execution and contains only the entries fetched by the queries.

\subsection{Database operations} \label{dataOper}
In this subsection, we describe basic database operations that affect the merge index.

\begin{algorithm}[!b]
Entry $result$[\ ] := $\emptyset$ \\
Entry $toInsert$[\ ] := $\emptyset$ \\

\ForEach {Key $k \in keys$} {

	Seek Entry $e$ with key $k$ in $ind$ \\
	\If {$e$ exists in $ind$} {
		$result$ := $result$ $\cup$ $e$ \\
	}
	\If {$k$ not in $InsertionJournal$ and
$k$ not in $DeletionJournal$}
{
		\tcp{The selection is made using partition attributes: min and max}
		Select Partition $par[]$ containing $k$ \\
		\ForEach {Partition $p \in par$} {
			Seek Entry $e$ with key $k$ in $p$ \\
			\If {$e$ exists in $p$} {
				$toInsert$ := $toInsert$ $\cup$ $e$ \\
				$result$ := $result$ $\cup$ $e$ \\
			}
		}
	}
}

\tcp{Adding entries from toInsert into the merge index starting from the root}

bulkload($root(ind)$, \ $toInsert$) \\

\tcp{Journal and partition set updating}
Pair $p$ := $(min(keys), \  max(keys))$ \\
updateInsertionJournal($p$) \\
updatePartitionSet() \\

\Return $result$

\caption{search(input: Index $ind$, Key $keys$[\ ])}
\label{pseudo:Search}
\end{algorithm}

\subsubsection{Search}

A search method (Algorithm \ref{pseudo:Search}) takes the merge index and a key range as an input. It returns the entries whose keys are inside the key range. The requested entries may be fetched either from the partition or from the merge index directly. The algorithm works as follows. If the key is found in the merge index (line 4), it can be added to the result set immediately (line 6). The condition (line 7) checks if the required key has not been copied from the partition to the merge index and has not been deleted from the partition. If it is true, the partitions which may hold the required entry are determined using their $min$ and $max$ attributes (line 8). If the required entry is found in the partition, it is added to $toInsert$ array (line 12) and to the result set (line 13). After finding all the keys, the entries are inserted into the merge index (line 14). To do this, the bulkload described in the next section is invoked.
At the end of the algorithm, the insertion journal and the partition set are updated (lines 16 to 17). The algorithm returns all entries of the required key range and modifies the merge index as a side effect.

\subsubsection{Insertion}

The insert operation is invoked when a new record must be added to the database. In our method, the insertion is made directly to the merge index without changing any partition.

\subsubsection{Deletion}

The delete operation is executed when the data must be removed from the database.
The data may be deleted either from the merge index or from the partition. In the first case, the index is modified in a way specific to the index type.
Typically, it is accomplished by inserting an entry marked as "ToDelete" into the merge index and then removing this entry during index reorganization.
In the second case, the entry is simply deleted from the partition. This fact is stored in the deletion journal. It may be a situation that an entry has been copied from the partition to the merge index and then removed from the index. In this case, the entry is stored in the deletion journal as well.

\subsubsection{Update}

The update operation is executed when a modification of the record occurs. In this situation, the old version of the record is deleted and the new one is inserted into the database.

\subsection{Crash recovery} \label{crash}

This subsection sketches a problem of crash recovery. We do not consider transaction issues because they are specific for the index type and are out of the scope of this paper. We concentrate only on recovering the data structures stored in DRAM:
\begin{itemize}
\item \textit{Insertion Journal.} It can be recovered by traversing all the keys in the index and creating key ranges from them. For example, if the crash happens after the search $a-h$ (see Figure \ref{fig:figure31}), index traversing creates $\langle (a-f), (i-i), (k-m) \rangle$ in the insertion journal. Although we lose some information about query ranges processed in the past (e.g. $(f-m)$), we still have correct information about real values inside the index.

\item \textit{Partition Set.} To recover the partition set, we must pass every partition. For each entry in the partition, we check whether the entry is written in the deletion journal. If so, we don't consider it for \textit{min} and \textit{max} value calculation.
\item \textit{Entry Log.} An entry can be added to the B+tree in DRAM but not merged with the main index yet. In this case, it would be lost in the system crash. To prevent it, we use the entry log stored on PCM similar to \cite{DBLP:journals/acta/ONeilCGO96}.
\end{itemize}

\section{Index choosing} \label{index}

The proposed framework can work with any index type. However, to exploit the PCM characteristics, we adapt and improve DPTree proposed in \cite{DBLP:journals/pvldb/ZhouSCHC19}. The core idea of DPTree is to batch multiple writes in DRAM persistently and, then, merge them into the main index on PCM. The index consists of two components (see Figure \ref{fig:figure4}). The first component is the B+ tree in DRAM where the data from operations like insert, delete, and update are collected. The second component is the main index, which is stored partly in RAM and partly in PCM. If the B+tree extends the fixed threshold, its entries are merged with the entries of the main index. In this way, index modification is performed in batches which drastically amortizes persistence overhead. Worth mentioning is the fact that a leaf node level is stored persistently on PCM and all inner nodes are inside DRAM. Thus, the inner node's reorganization cost is avoided.

We did some improvements in DPTree. We observed that adaptive merging can insert many batches with entries sorted by key. As a result, the sorted entries once inserted into a leaf node remain intact for a long time. On the other hand, database modifications can add some new entries into the sorted leaf nodes. In this case, it is better to insert such entries at the end of the node in an unsorted fashion than to reorganize the entire node.
To accomplish this, we divide a node into two sections: sorted and unsorted. To find an entry in the sorted section, we utilize a binary search. Additionally, each node has a bitmap that indicates if the entry is valid or not. When the entry is deleted from the node, it is not removed immediately from PCM but marked as invalid in the bitmap.
The size of both sections is aligned with the multiplication of the cache line size. In this way, the number of cache misses is reduced during linear searching.

The node size is an important parameter that affects the performance of the tree. Previous works suggest that a tree node size should be equal to the size of a few cache lines. For modern computers, the cache line size is usually 32, 64, or 128 bytes. We fix the cache line as 64 bytes for all the methods. Consequently, the node size of the BB+tree in PAM is 512 bytes.

\medskip
In summary, our index, named BB+tree, has the following features:
\begin{itemize}
\item It contains the DRAM B+tree where the data are collected.
\item It contains the main index with leaf nodes on PCM and inner nodes in DRAM.
\item A leaf node has two sections: sorted and unsorted.
\item The bulkload operation merges the batches stored in B+tree with the entries in the main index.
\end{itemize}

\subsection{Merge index modification algorithms}

As we mentioned before, two types of database operations can affect the merge index: searching and modification. The data obtained from these operations are first collected in the B+tree in DRAM. When the B+tree is full, the bulkload is invoked.

In this subsection, we present two algorithms for index modification. The first algorithm describes bulkload. The second one reflects the inserting of the entry into the main index.

\begin{algorithm}[!h]
\caption{bulkload(input: Node $node$, Entry $entries$[\ ])}
\If{length($entries$) = 0} {
	\Return
}
\If {isLeaf($node$) = $TRUE$} {
	insert($node$, \ $entries$) \\
	\Return
}
Key $lastMax$ := $-\infty$ \\

\tcp{Go down to the children nodes with the subset of entries with keys between $minKey$ and $node$.keys[i]}
\For{$i \gets 1$ to $FANOUT$} {
	Key $minKey$ := $lastMax$  \\
	\tcp{node.keys[FANOUT] is always equal to $\infty$}
	Key $maxKey$ := $node$.keys[i] \\
	$lastMax$ := $maxKey$ \\
	Entry $slice$[ \ ] := sliceMake($entries$, \ $minKey$, \ $maxKey$) \\
	bulkload($node$.child[i], \ $slice$) \\
}
\label{pseudo:Bulkload1}
\end{algorithm}

\vspace*{2mm}

\begin{algorithm}[!h]
\caption{insert(input: Node $node$, Entry $entries$[\ ])}
Node n = node

\ForEach {Entry $e \in entries$} {

		\If{$nmbValidEntries(n) <  FANOUT/2$} {
			n = MergeNodes() \\
			UpdateInnerNodes()	\\		
		}

		\If{$nmbValidEntries(n) > FANOUT$} {
			n = SplitNode() \\
			UpdateInnerNodes() \\
		}

	\If{$e$.$ToDelete$ = $TRUE$} {
	 	delete(n,e) \\
	}
	\Else{

		\If{$existsGapInUnsorted(n)$ = $TRUE$}  {
		insertIntoUnsortedGap(n,e) \\
		}
 	\ElseIf {$unsortedNotFull(n)$ = $TRUE$} {
		insertIntoUnsorted(n,e) \\
		}
	\ElseIf{$existsGapInSorted(n)$ = $TRUE$} {
		insertIntoSortedGap(n,e) \\
		}
}

}
\label{pseudo:Insert}
\end{algorithm}

Algorithm 2 describes bulkload as a dominant operation in adaptive merging. The method inserts many entries into the main index in batches. It starts in the root node and works recursively. The procedure takes a node and an entry set as an input. As the root and inner nodes have pointers to the child nodes, the input entry set is split according to the key ranges of the child nodes. It is reflected in lines 6 to 12. When the leaf level is reached, the appropriate entries are added to the leaf (lines 3 to 5). For this, algorithm 3 is responsible.

Algorithm 3 inserts the entries into the leaf nodes. When the inserted entry is marked as "ToDelete", it is removed from the leaf node of the main index. It is reflected in lines 9 and 10.

\begin{figure}[!b]
\centering
\includegraphics[width=0.45\textwidth]{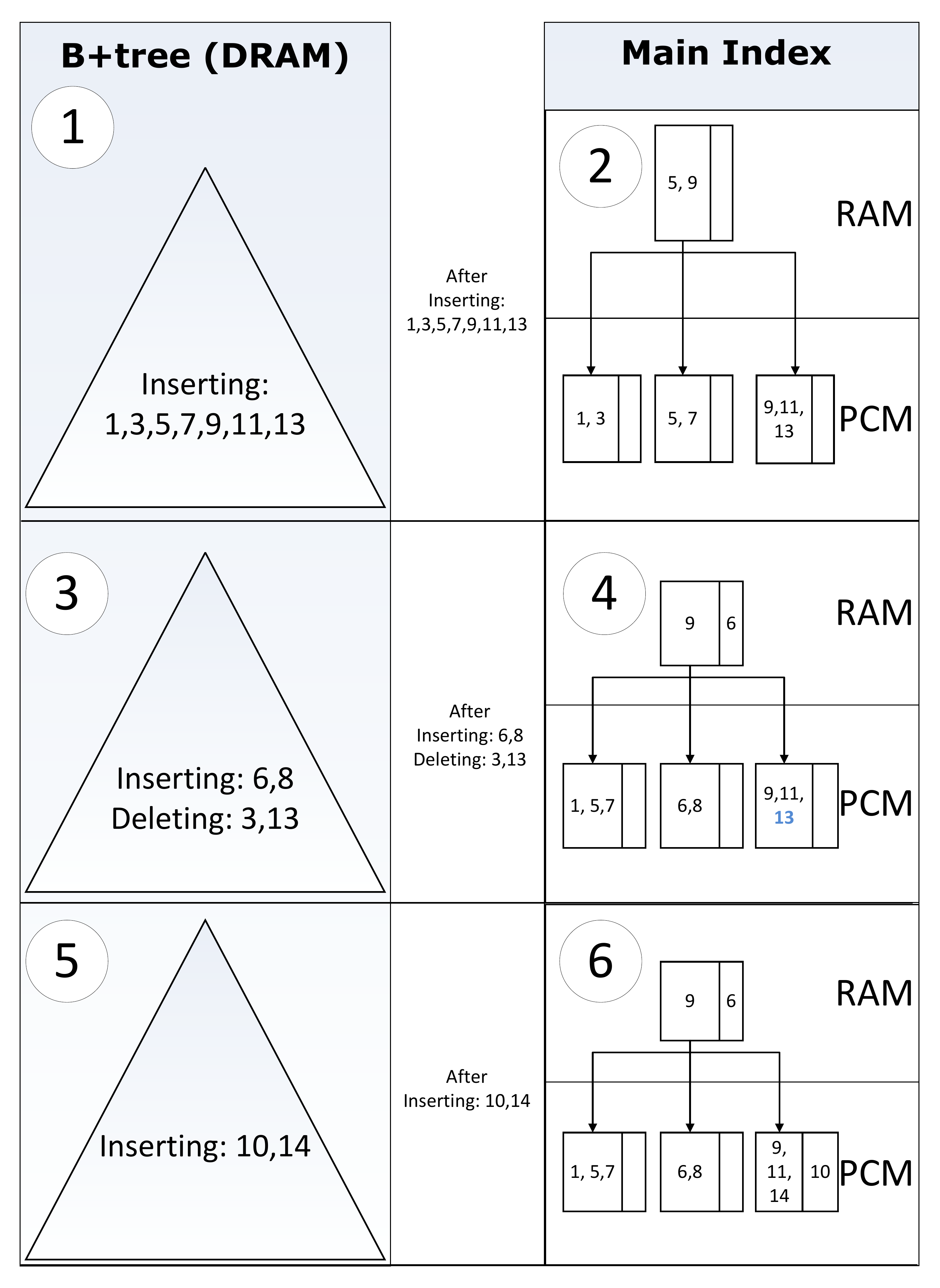}\vspace*{-2mm}
\caption{Bulkload example}
\label{fig:figure4}
\end{figure}

When some entry has been removed from the leaf and the leaves have not been reorganized yet, memory gaps are created. They can exist either in the sorted or in the unsorted section of the node. When the normal entry is inserted into the leaf node, the algorithm checks the gaps in the unsorted section. If it finds them, the entry is inserted into the gap (lines 12 and 13). If not, the entry is inserted into the free place in the unsorted section (lines 14 and 15). When the unsorted section is full, the algorithm seeks the gap in the sorted section of the node (lines 16 and 17). When there is no place in the requested leaf node, a new one must be created. In the other case,
the index must be reorganized in the following way.
If the number of valid entries in the node is less than $fanout/2$, the nodes are merged (lines 3 to 5). On the other hand, if the number of valid entries in the node is more than $fanout$, the node is split (lines 6 to 8). In both cases, the inner nodes up to the root node can be reorganized (lines 5 and 8).

Figure \ref{fig:figure4} shows an example of the index modification. We assume that the fanout of the main index is equal to 4. Each node of the main index is divided into sorted (left) and unsorted (right) sections. The example consists of six steps. In the first step, a batch containing seven new entries is added to the B+ tree in DRAM. After that (step 2), the main index is created and the entries are inserted into it using the bulkload method. Please note that the main index has three leaf nodes on PCM and one inner node in DRAM. As it is the first batch, all entries inside the nodes remain sorted. In step 3, a batch consisting of two entries to insert and two entries to delete is added to the B+tree. The merging (step 4) is performed as follows. Algorithm 2 is invoked for the root node and the entry set. The entry set is split into three subsets because the root node has three children. This is reflected in the recursive section of the bulkload method (lines 7 to 12). After that, the insertion to the leaf nodes is performed in the following way.
Entry with key 3 is removed from the leftmost leaf node. As the number of entries in the node must be at least two, this node must be merged with the middle node. Now, the newly created node contains the entries: 1,5,7. After that, entries 6 and 8 are inserted to this node. Since the node has five entries, a node split is needed. As a consequence, two nodes containing the entries: 1,5,7, and 6,8 are created.
At last, entry 13 is marked as deleted. This fact is denoted using the light font. The leaf node update results in the reorganization of the inner nodes. Thus, the root node is changed: key 5 is removed and key 6 is added to the unsorted section of the root node.
In the last batch, two entries: 10 and 14 are added to the B+ tree (step 5). Both entries should be inserted into the rightmost leaf node. Entry 10 is inserted into the unsorted section of the node and entry 14 fills the gap formed by removing entry 13 in the sorted section (step 6).

\section{Performance evaluation} \label{exp}

The section consists of two parts. In the first one, we describe the  implementation of the PCM simulator. The second part presents some experiments which confirm the effectiveness of the proposed methods.

\subsection{Implementation issues}

\subsubsection{PCM simulator}

The core of our system is a PCM simulator.
The implementation is written in C and compiled with gcc 9.3. We simulate PCM in the same way as proposed in \cite{RethinkingDatabase}.
First, we model data comparison writes for PCM writes. We write a 64B cache line to PCM and compare this line with the original one to compute the number of modified bits and the number of modified words.
Second, we model eight PCM memory ranks with parallel access.
At last, we model the details of cache line write back operations.
Since we have eight parallel 8B ranks, we can do a flush cache (64B) as fast as one 8B write. In modern PCM, most of the work is done by the controller. However, it depends on the software developer how the entries in the write queue are arranged.

\medskip
Typically, the controller manages 32 or 64-entry FIFO write queue.
Even if the controller can do writing in the background, it is fully stressed during write-intensive workloads (the write queue is almost always full). That is why, we favor perfect cache utilization, i.e. writes that need to change 64B are better than several writes changing 8B sequentially one after another.
To accomplish this utilization, we introduce a new way to calculate fill factor for each index. Typically, the bulkload creates the node with a fixed 70 - 80\% fill factor. In our method, we use 80\% with alignment to 64B. We initialize a node with an unsorted area aligned to several cache lines. In this way, the linear search works on full cache lines.

\medskip
In the simulator, we fixed the following PCM parameters:
\begin{itemize}
\item Page size: 64B
\item Page read latency: 50 ns
\item Page write latency: 1 $\mu$s
\item Write bandwidth: 64 MB$\backslash$s per die
\end{itemize}

\subsection{Experiments}

We divide the experiments into three parts.
The first part compares PCM-optimized indexes. We consider three candidate index types and choose the best for our purpose.
In the second part, we compare all implemented methods of adaptive merging under different workloads. Each experiment is carried out until the merge index is completed.
In the third part, we place our method in a dynamic database environment where database modifications interleave with search queries.

The experiments were conducted on Intel Core i7-6700k 4.0Ghz (8CPUs) equipped with 32 GB RAM.
For each experiment, we use a set of 100 million entries. The size of one entry is 16 bytes making the total size of the dataset equal to 1.6GB.

\subsubsection{Index comparison}

In this section, we experimentally compare different index types.  Some of them were already mentioned in the previous subsections. Following PCM tendency, the inner nodes of all indexes are held in DRAM. The goal of the experiment is to justify that BB+tree is the most suitable index for adaptive merging on PCM. We do not consider the indexes like CB+tree (see \cite{edbtLiJYWY16}) with inner nodes stored on PCM since their reorganization cost is high.

\medskip
We consider the following index types:
\begin{itemize}
\item UB+tree - B+tree with unsorted leaves.
\item SB+tree - B+tree with two section nodes: sorted and unsorted.
\item BB+tree - buffered B+tree with two section nodes: sorted and unsorted.
\end{itemize}

In the experiments, we used the following workloads:
\begin{itemize}
\itemsep=0.9pt
\item \textit{Write intensive workload}: 10 batches, each batch contains 40000 inserts, 40000 deletes, and 20000 point searches.
\item \textit{Read intensive workload}: 10 batches, each batch contains 10000 inserts, 10000 deletes, and 80000 point searches.
\item \textit{Balanced workload}: 10 batches, each batch contains 25000 inserts, 25000 deletes, and 50000 point searches.
\end{itemize}

We estimate index performance without executing an adaptive merging process. Like in the traditional indexing, we first create and fill each index with 100 million entries using bulkload. As a consequence, we obtain the index with a completely filled sorted section and an empty unsorted section. Then, each of the above mentioned workloads is applied. It contains batches with a different number of operation types.

\begin{figure*}[!h]
\vspace*{2mm}
\begin{minipage}[b]{0.45\linewidth}
\centering
\includegraphics[width=\linewidth]{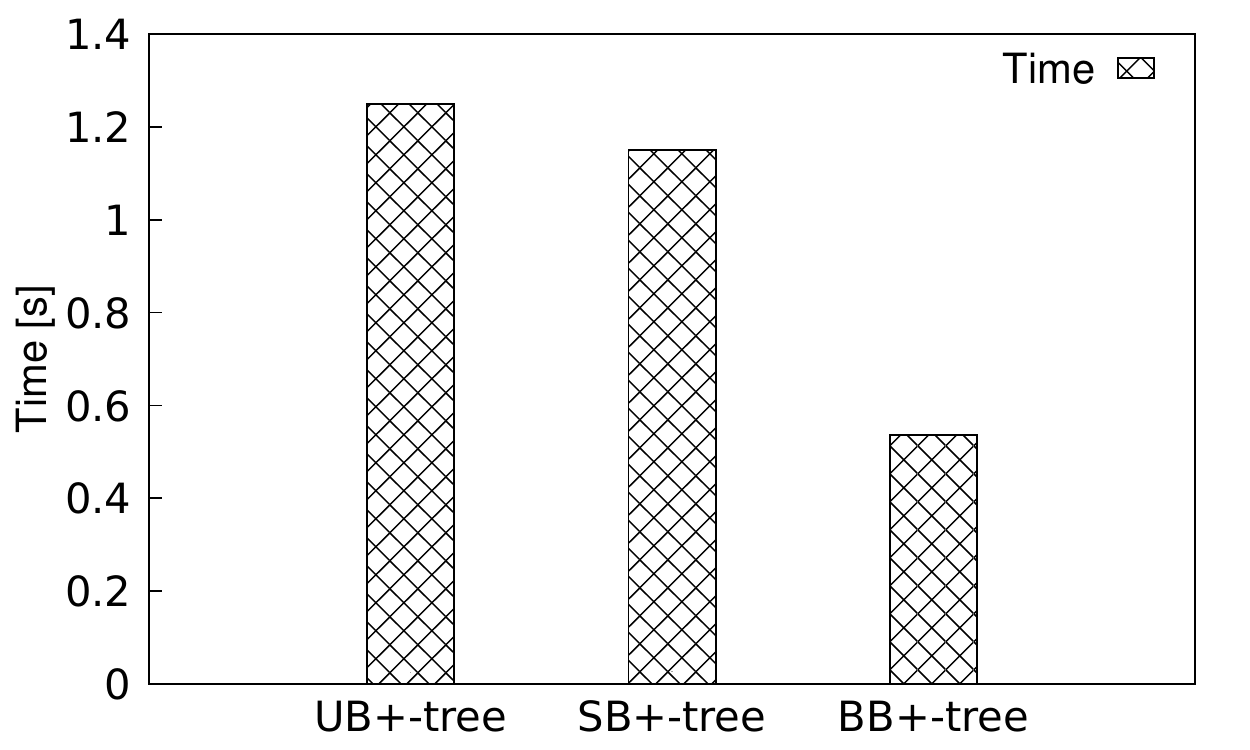}\vspace*{-2mm}\quad
        \caption{Balanced workload}
        \label{fig:ex2_index_balanced}
\end{minipage}
\hfill
\begin{minipage}[b]{0.45\linewidth}
\centering
\includegraphics[width=\linewidth]{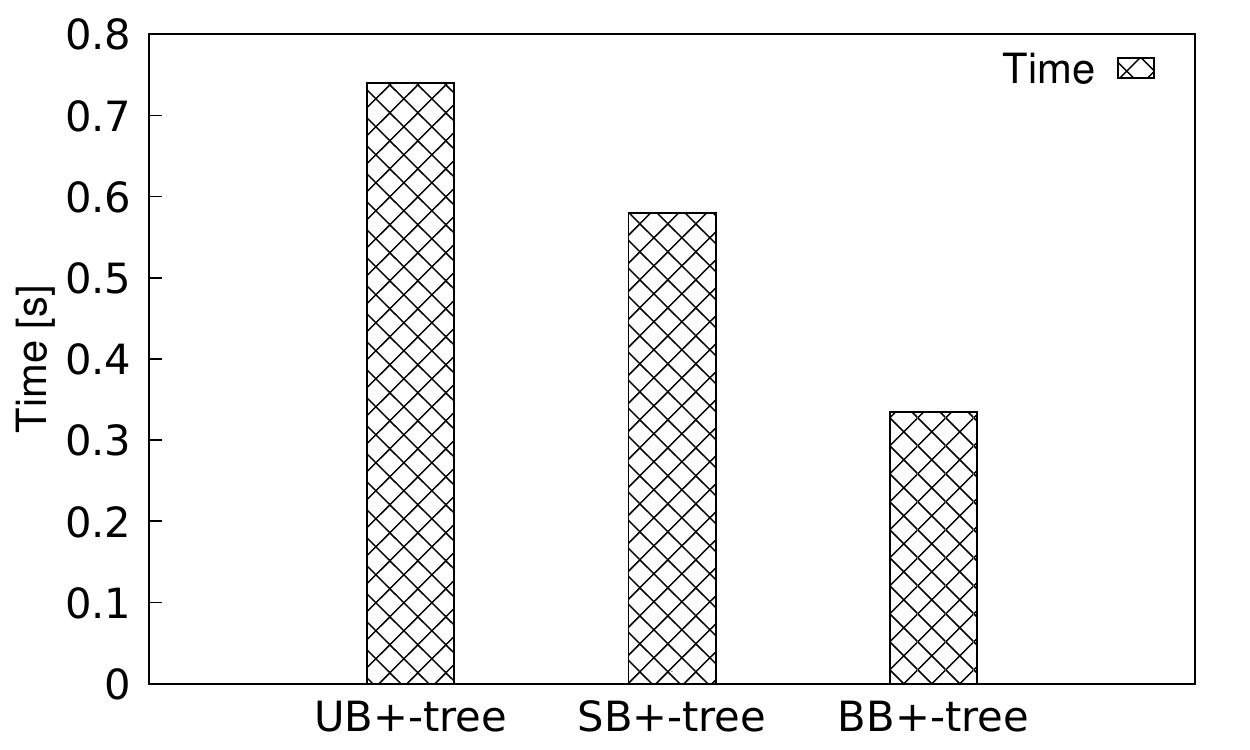}\vspace*{-2mm}\quad
        \caption{Read intensive workload}
        \label{fig:ex2_index_read_intensive}
\end{minipage}\vspace*{-2mm}
\end{figure*}

\begin{figure*}[!h]
\begin{minipage}[b]{0.45\linewidth}
\centering
\includegraphics[width=\linewidth]{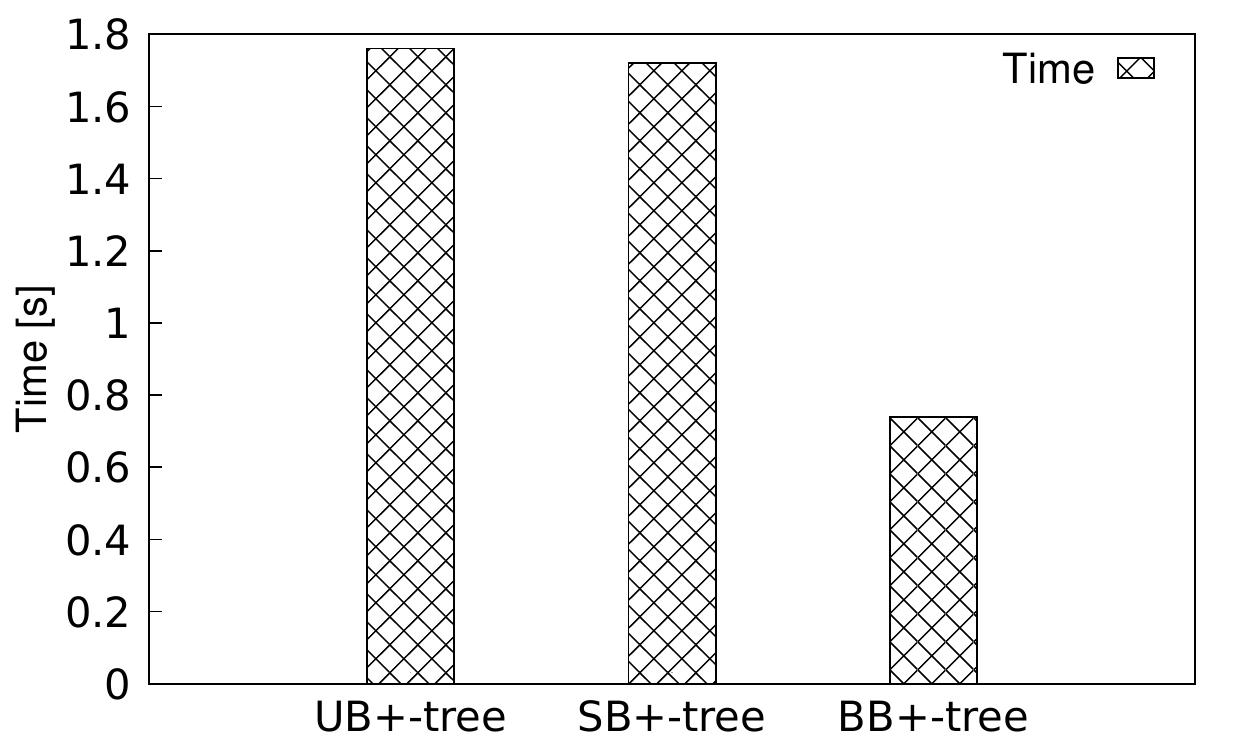}\vspace*{-2mm}\quad
        \caption{Write intensive workload}
        \label{fig:ex2_index_write_intensive}
\end{minipage}
\hfill
\begin{minipage}[b]{0.45\linewidth}
\centering
\includegraphics[width=\linewidth]{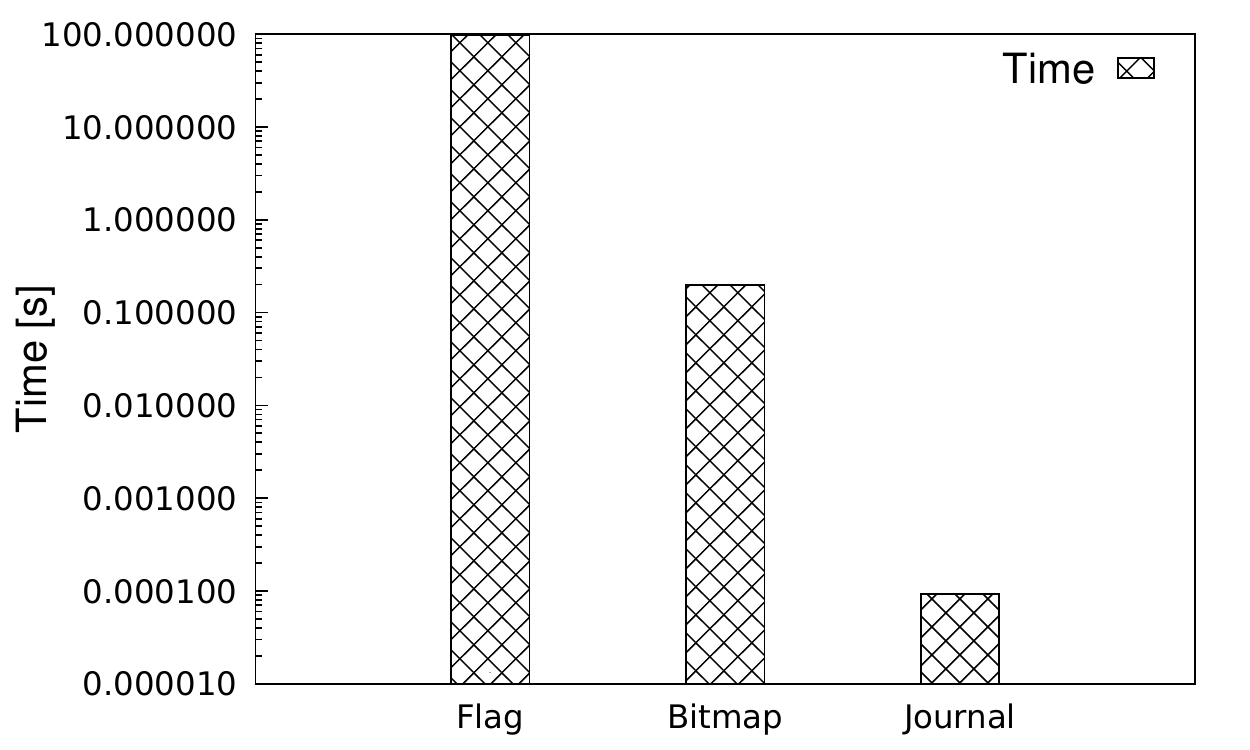}\vspace*{-2mm}\quad
        \caption{Invalidation}
        \label{fig:ex1_invalidation}
\end{minipage}
\end{figure*}

As we can observe (see Figures \ref{fig:ex2_index_balanced}, \ref{fig:ex2_index_read_intensive} and \ref{fig:ex2_index_write_intensive}), introducing sorted and unsorted sections of the node (SB+tree) slightly decreases the elapsed time for the balanced and write-intensive workload. The read-intensive workload shows better performance since it exploits the binary search in the sorted section of the node.
The application of the buffered index with sorted and unsorted sections (BB+tree) significantly improves the performance of the whole system. It is because the data are collected and merged with entries of the main index in batches.
In this way, we experimentally prove that BB+tree is a strong candidate for adaptive merging on PCM.

\subsubsection{Adaptive merging implementations} \label{Impl}

In this subsection, we compare three different adaptive merging implementations. All of them use different index types. Following the current tendency, the inner nodes of each index are stored not on PCM but in DRAM. In the initial step of each implementation, all merge index entries are sorted and inserted into partitions. Thus, each partition holds the entries sorted by $key$.

\medskip
We consider the following methods:
\begin{itemize}
\item \textit{Original adaptive merging (AM).} The implementation is based on the method proposed for traditional block memory types \cite{Idreos:2009:STR:1559845.1559878}. In this method, a partitioned B+tree is built over the partitions and each of its nodes corresponds to one partition. As the merge index entries are sorted only in a partition's scope, the entries with the same keys can be stored in many partitions. Thus, an entry of the partitioned B+ tree may point to many partitions. The approach is not suitable for PCM, since it shifts the data after each modification like in the traditional B+-tree. To make the implementation PCM friendly, we did some improvements. We created a memory pool for storing entries for deletion. The memory pool is useful for late materialization. This technique is used when the modification is not immediately reflected in the index but is postponed until the memory pool is full. Thus, if a range query is performed, the memory pool is scanned. The deleted merge index entries are removed from the result provided that they fulfill the search condition. Utilizing the memory pool drastically improves the method for PCM.

\item \textit{Extended Adaptive Merging (eAM).}
The goal of this method is to adjust the standard adaptive merging (AM) to PCM characteristics. We observed that the operations like insert, delete, and search cause node modifications in the partitioned B+tree. Such a situation has a negative impact on PCM performance and increases the number of memory writes. To prevent it, we replaced the partitioned B+tree by UB+tree with a bitmap. The entries inside the leaf node are unsorted to avoid unnecessary memory overhead. Each entry is connected with one bit of the bitmap. If the entry is valid, the bit is set to 1. Otherwise, the bit is set to 0. In this way, any index modification results in changing one bit instead of reorganizing the entire tree node. Moreover, each partition is equipped with a bitmap as well. The bit in the bitmap indicates whether the entry is already moved from the partition to the index.

\item \textit{PCM Adaptive Merging (PAM).} The implementation is based on the framework proposed in this paper (see subsection \ref{frm}). As the merge index, we use BB+tree described in more detail in section \ref{index}
\end{itemize}

This experimental series considers index convergence. It means that each experiment is conducted until all entries from the database table are inside the index.
To do this, we use range queries with selectivity equal to 5\% and data patterns described below. By $selectivity$, we denote which fraction of records is fetched by the query. We assume that the database is not modified during the adaptive merging. A combination of adaptive merging and database modification is considered in the next subsection.

\eject
For these experiments, we use the following data patterns:
\begin{itemize}
\item Random pattern - we draw a minimal key ($min$) with a uniform distribution. Then, we select the records starting with this key so that the fixed selectivity is reached. Thus, we obtain the range $[min, min + rows]$ where $rows = totalRows * selectivity$.

\item Sequential pattern -
we create the sequential access pattern as follows (see \cite{SekPattern}): starting from the beginning of the value domain, the queried range is shifted for each query by half of its size towards the end of the domain. When the end is reached, the query range restarts from the beginning. The position to begin is randomly set in the first 0.01\% of the domain. In this way, the repetition of the same sequence in subsequent rounds is avoided.

\item New keys pattern - each query asks for non-index data.
\end{itemize}

In the first experiment (Figure \ref{fig:ex1_invalidation}), we compare three invalidation methods for the data fetched from the partitions:
\begin{itemize}
\item Flag - the invalid data has an invalidation bit set to \textit{false}.
\item Bitmap - each partition contains its own bitmap where data invalidation is registered (see \cite{Chi:2014:MBT:2627369.2627630}).
\item Journal - the invalidation technique applied in PAM that uses the journal to indicate valid data.
\end{itemize}

We execute AM with the random pattern and selectivity 5\% until the full index is created. Figure \ref{fig:ex1_invalidation} presents the total time of invalidation operations for three invalidation methods. The worst performance shows the flag invalidation method (Fig. \ref{fig:ex1_invalidation}). It comes from the fact that one byte must be updated after the invalidation of each data. For bitmap invalidation, each data of the partition is associated with a single bit.
Since selectivity is 5\%, a lot of entries are marked by the same query. Thus, the bitmap can fully utilize the cache line. As the cache line is 64B, we can mark 512 entries at the same time.
The best performance is notified by utilizing the journal.

\begin{figure*}[!h]
\begin{minipage}[b]{0.45\linewidth}
\centering
\includegraphics[width=\linewidth]{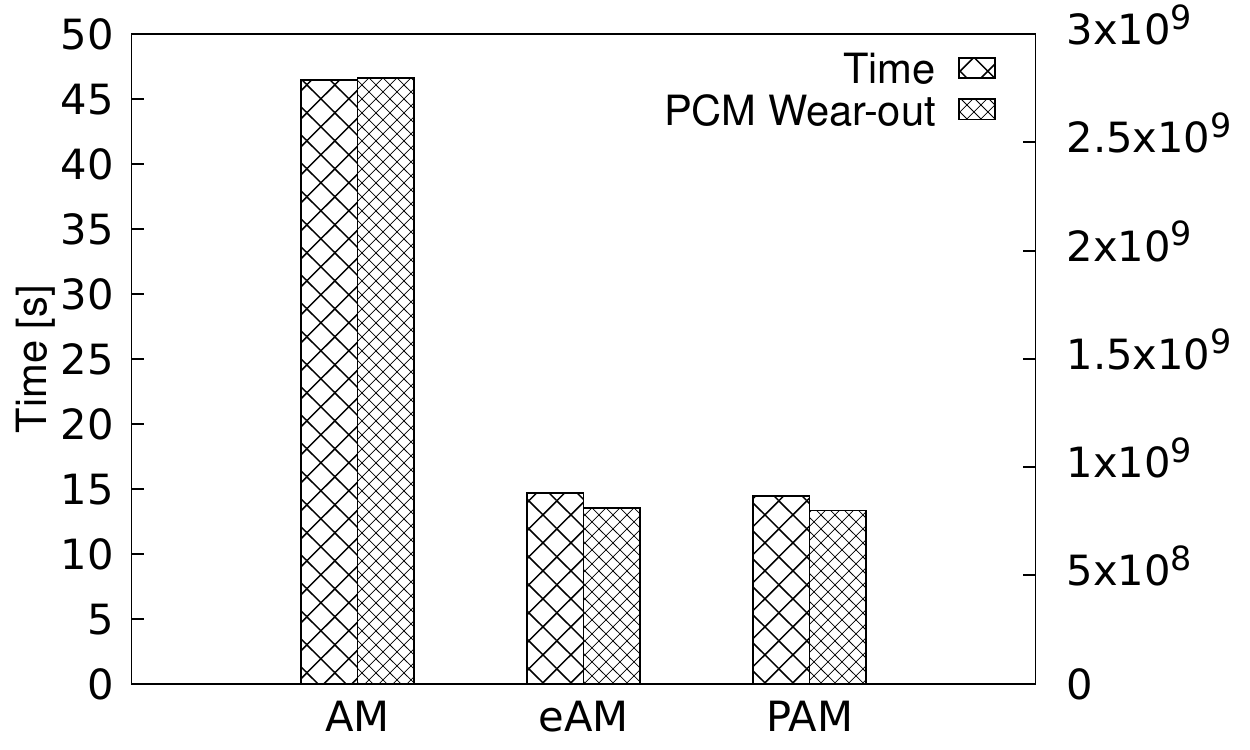}\quad
        \caption{Sequential total}
        \label{fig:ex3_seq_total}
\end{minipage}
\hfill
\begin{minipage}[b]{0.45\linewidth}
\centering
\includegraphics[width=\linewidth]{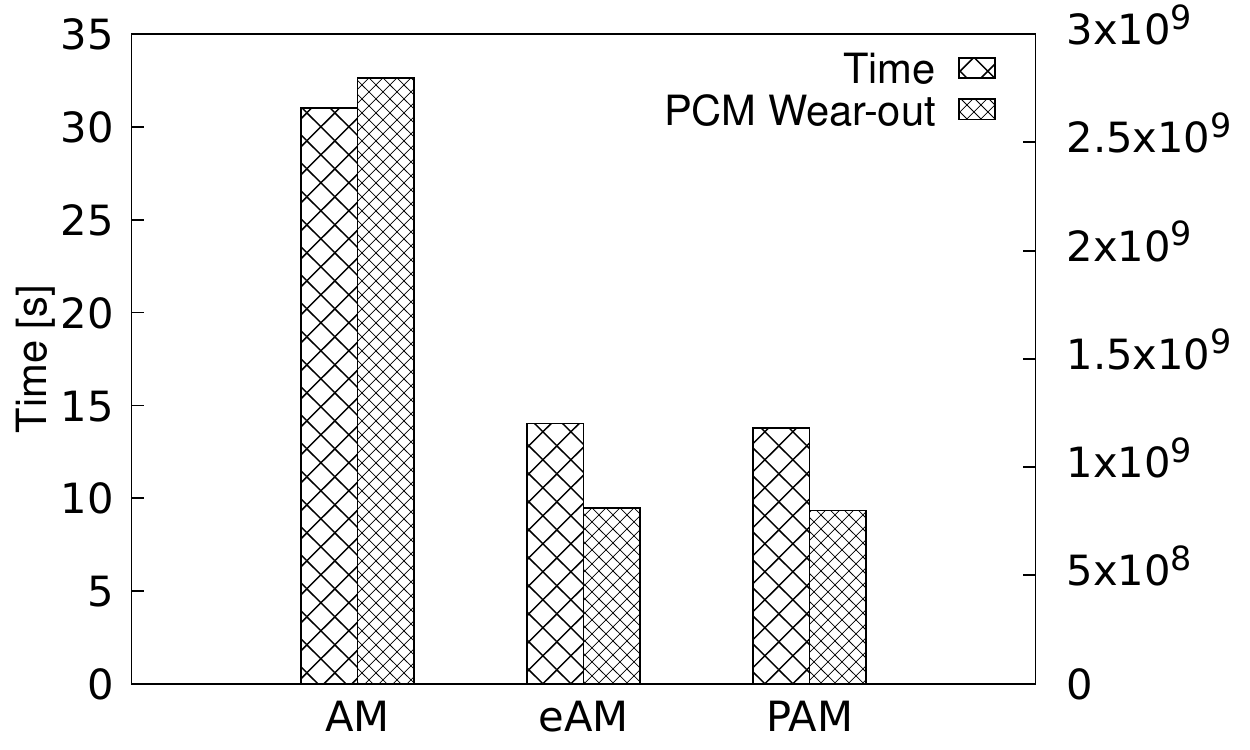}\quad
        \caption{New keys total}
        \label{fig:ex3_newkeys_total}
\end{minipage}
\end{figure*}

\begin{figure*}[!h]
\begin{minipage}[b]{0.48\linewidth}
\centering
\includegraphics[width=\linewidth]{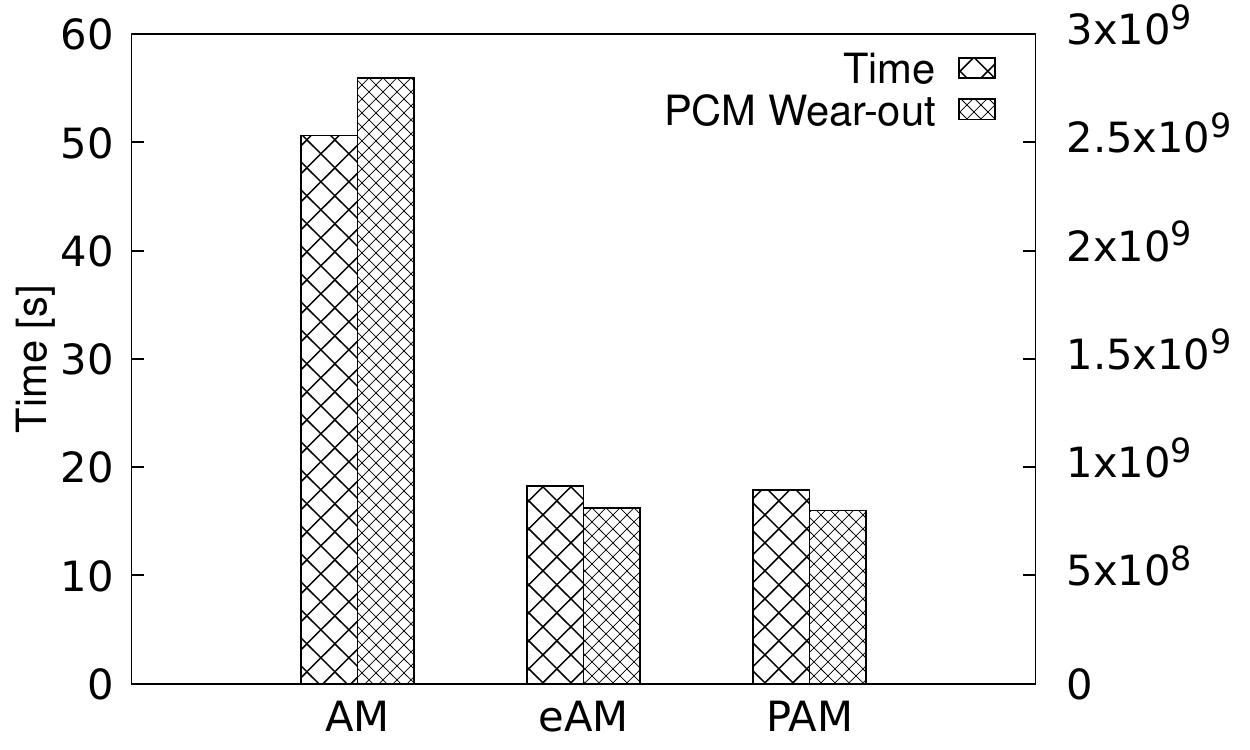}\quad
        \caption{Random total}
        \label{fig:ex3_random_total}
\end{minipage}
\hfill
\begin{minipage}[b]{0.48\linewidth}
\centering
\includegraphics[width=\linewidth]{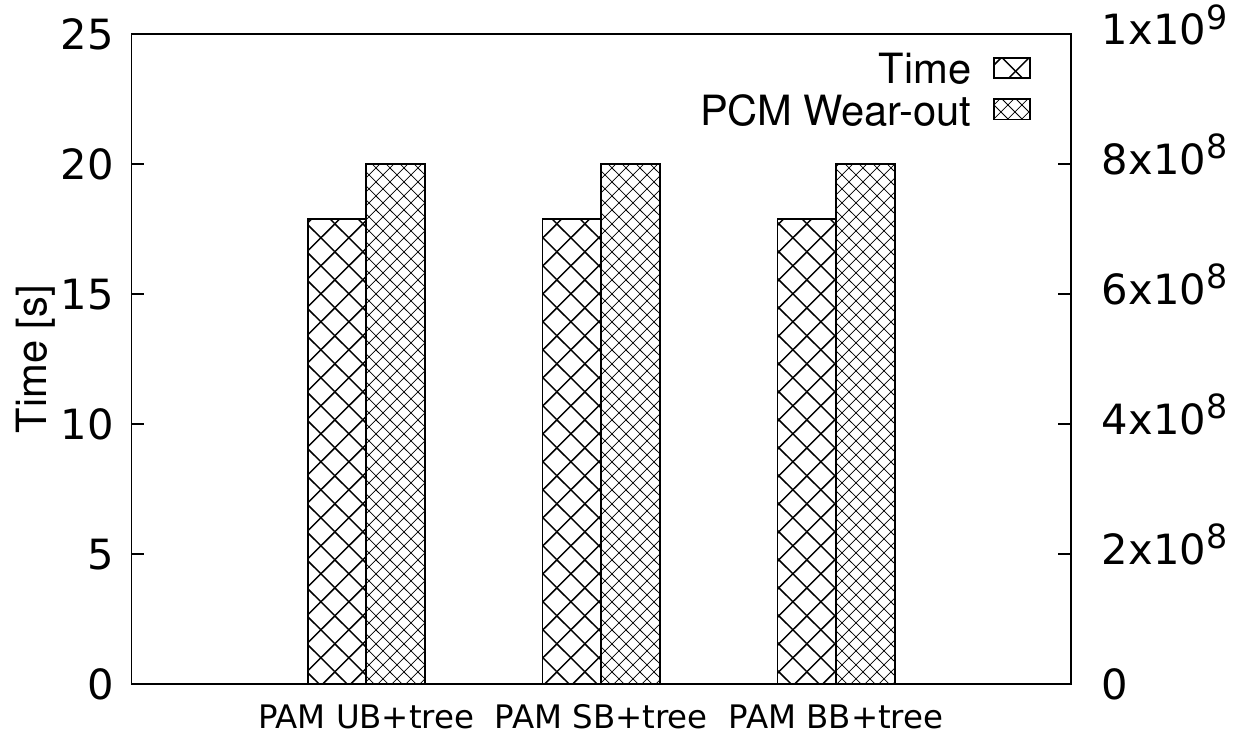}\quad
        \caption{Random total with different index types}
        \label{fig:ex3_1_random_total}
\end{minipage}
\end{figure*}

In the next experiment, we measure the elapsed time and PCM wear-out for the different data patterns (see Figures \ref{fig:ex3_seq_total}, \ref{fig:ex3_newkeys_total}  and \ref{fig:ex3_random_total}). As we can see, the traditional adaptive merging (AM) exhibits the worst performance. It comes from the fact that the method is not optimized for byte-addressable memory. Both methods: eAM and PAM show similar performance (the difference is about 1.2 seconds in favor of PAM). The same trend can be noticed for memory wear-out.
In the experiment presented in Figure \ref{fig:ex3_1_random_total}, we applied the random pattern for PAM with different index types. As we can see, the index type does not affect PAM in the case when the adaptive merging without database modifications is considered.

\subsubsection{Update cracking database}

In this subsection, we show that the nice performance properties of adaptive merging can be maintained in a dynamic environment where database modifications interleave with search queries. We carry out the experiments for the workloads with database modifications. Each workload consists of a fixed number of batches, whereas each batch is a collection of delete, insert and range searches with selectivity from 1\% to 5\%.
Using the different workloads, we check when the framework and when index type affect the performance of adaptive merging. For that, we compare eAM and PAM with the different index types inside. Additionally, we measure the impact of search selectivity on the performance of the methods. For each workload, we perform five experiments for selectivity from 1\% to 5\% in the range search.
We do not show the results for AM, since its performance is very bad. It comes from the fact that AM is not optimized for database modification on PCM.

\medskip
We use the following workload types:
\begin{itemize}
\item \textit{WorkloadA}: 100 batches, each batch contains: 5 inserts, 5 deletes, and 10 range searches.
\item \textit{WorkloadB}: 5 batches, each batch contains: 100000 inserts, 100000 deletes, and 5 range searches.
\item \textit{WorkloadC}: 10 batches, each batch contains: 100000000 inserts, 100000 deletes, and 20 range searches.
\item \textit{WorkloadD}: 10 batches, each batch contains: 10000000 inserts, 10000 deletes, and 10 range searches.
\end{itemize}

\begin{figure*}[ht]
\begin{minipage}[b]{0.45\linewidth}
\centering
\includegraphics[width=\linewidth]{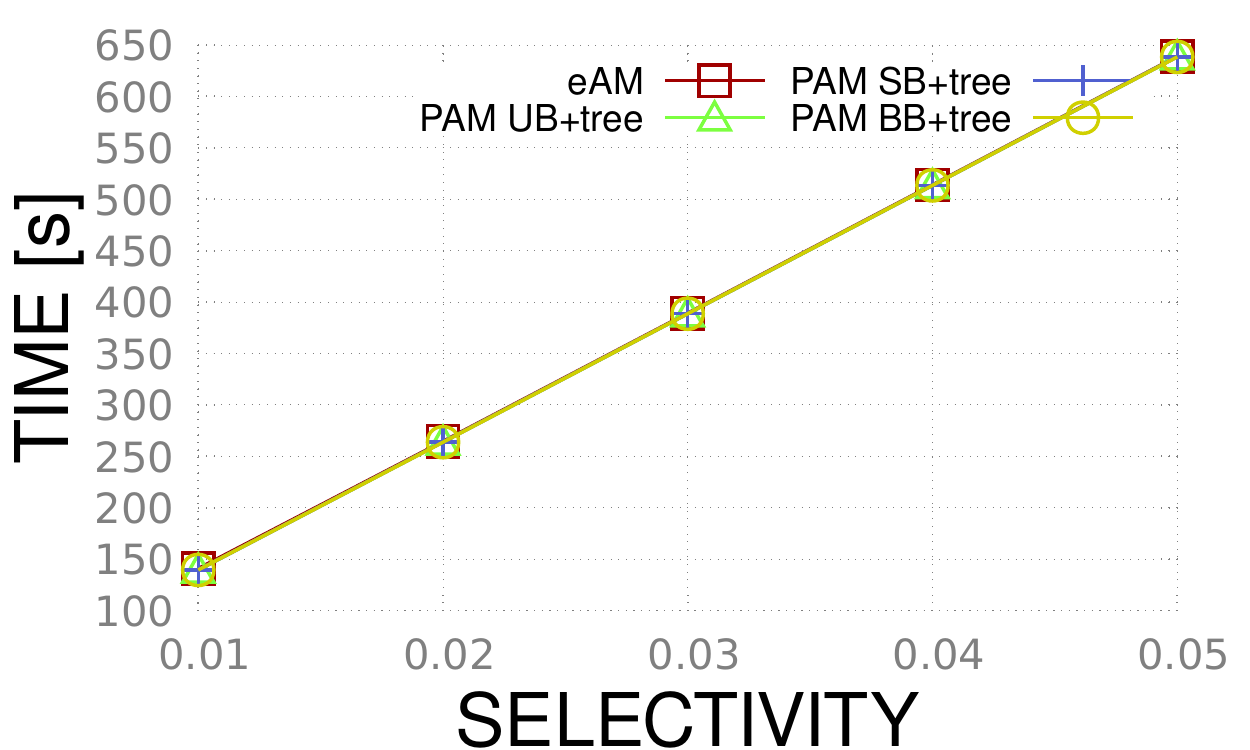}\vspace*{-2mm}\quad
        \caption{WorkloadA}
        \label{fig:ex4_stress1}
\end{minipage}
\hspace{0.5cm}
\begin{minipage}[b]{0.45\linewidth}
\centering
\includegraphics[width=\linewidth]{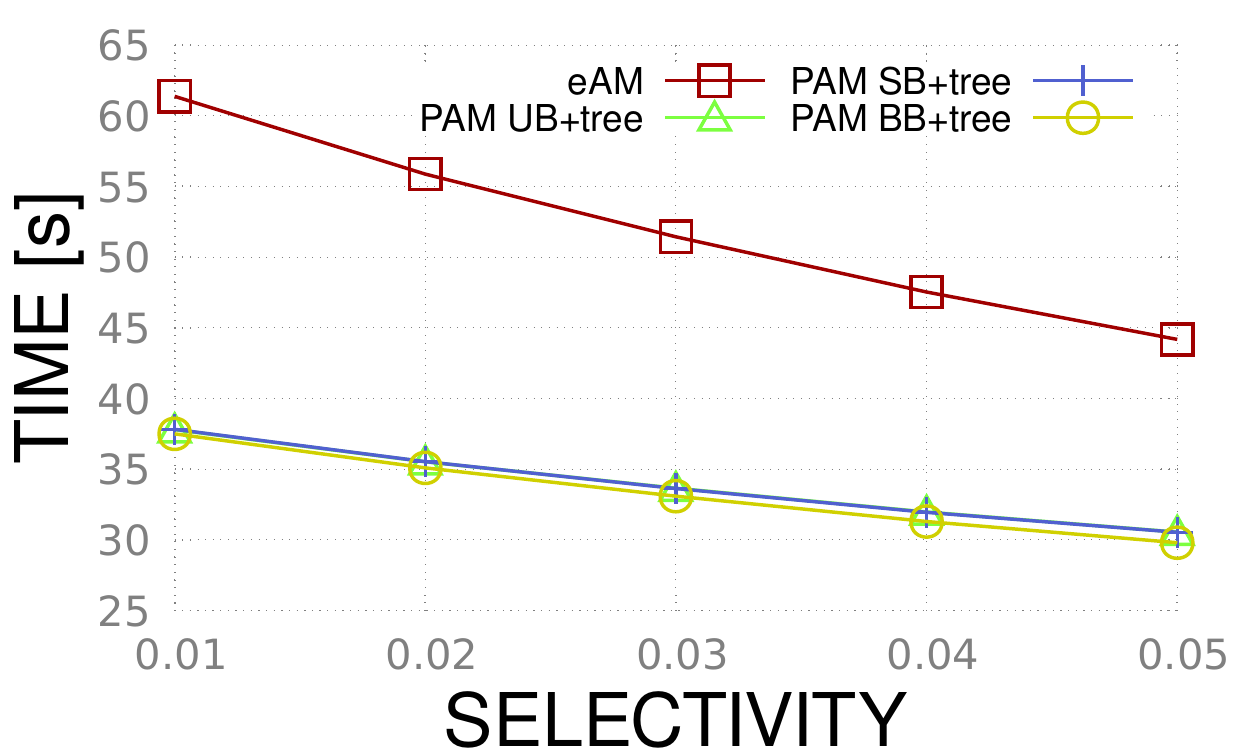}\vspace*{-2mm}\quad
        \caption{WorkloadB}
        \label{fig:ex4_stress2}
\end{minipage}
\end{figure*}

\begin{figure*}[ht]
\begin{minipage}[b]{0.48\linewidth}
\centering
\includegraphics[width=\linewidth]{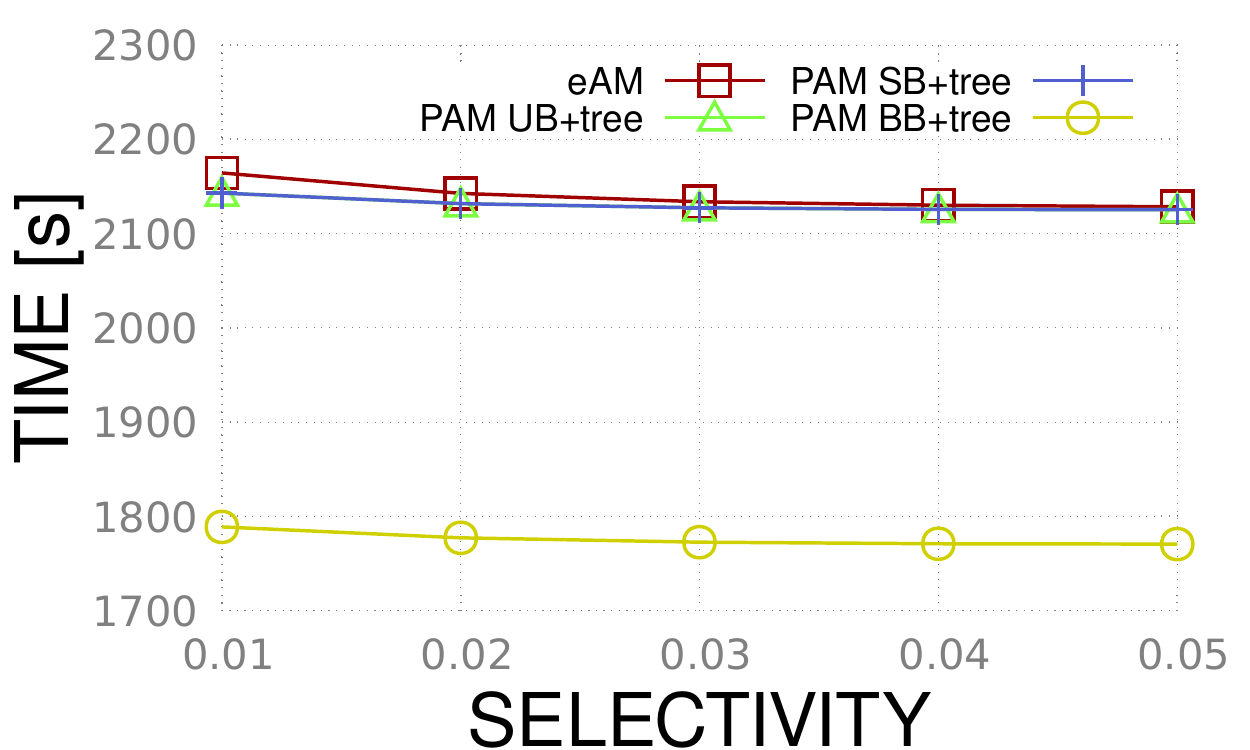}\vspace*{-2mm}\quad
            \caption{WorkloadC}
        \label{fig:ex4_stress3}
\end{minipage}
\hspace*{5mm}
\begin{minipage}[b]{0.48\linewidth}
\centering
\includegraphics[width=\linewidth]{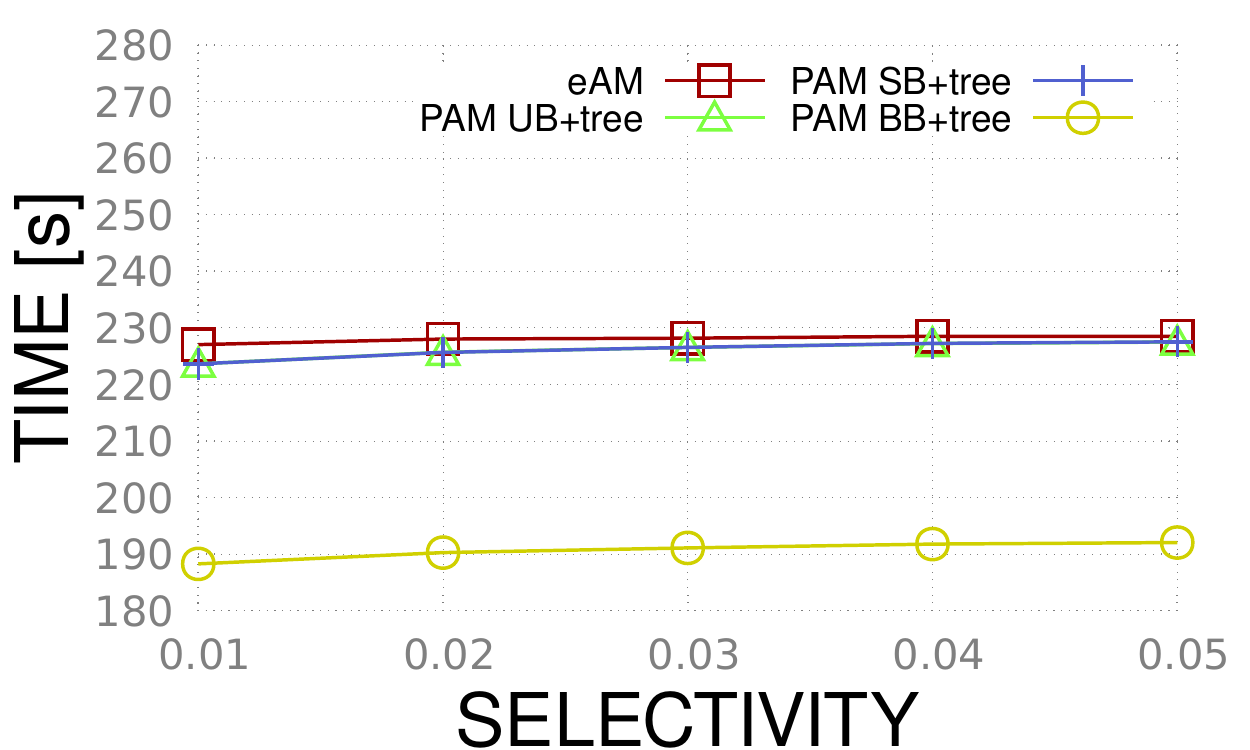}\vspace*{-2mm}\quad
        \caption{Random total with different index types}
        \label{fig:ex4_stress6}
\end{minipage}
\end{figure*}

Figure \ref{fig:ex4_stress1} presents the elapsed time for \textit{WorkloadA}. In this case, we need 5 to 10 (depending on selectivity) batches to complete the merge index. So, the framework works very short and the system hardly benefits from using it. As a result, the performance is similar for all methods: eAM and PAM with the different index types. The execution time does not depend on the index type, since the number of data operations is too small.
Additionally, we can observe that the execution time increases with increasing selectivity. When the selectivity grows, the range queries need more time to fetch the data.

\medskip
Figure \ref{fig:ex4_stress2} presents \textit{WorkloadB} where database modifications are dominant. Since the number of range searches is not large, we do not reach a state where the merge index is completed. When the selectivity increases, the range queries fetch more data. As a consequence, more data is inside the merge index and less data is stored in the partitions. Thus, the overall system performance decreases.
Moreover, PAM outperforms eAM. It is due to the optimization proposed in the framework. The use of the deletion and insertion journals significantly improves the system's performance. When the merge index is not completed, the framework can manage the partitions very efficiently.
Furthermore, selectivity affects cache utilization in eAM. As we mentioned before, each partition is equipped with a bitmap. A bit in the bitmap indicates whether the entry is already moved from the partition to the merge index.
When the selectivity is small, fewer data are fetched from the partitions and fewer bits in the bitmap are updated at the same moment. As a consequence, the number of cache misses is large.
This is not the case in PAM because it has the insertion journal instead of the bitmap.
We can also observe that all index types show the same performance. They behave similarly because the number of range searches is too small to fully exploit the advantages of using the two-section index.

\medskip
Figures \ref{fig:ex4_stress3} and \ref{fig:ex4_stress6} present the performance of \textit{WorkloadC} and \textit{WorkloadD}, respectively. In both cases, the framework works very short and after that, many database modifications are performed. As we can see, all methods except PAM with BB+tree look similar. Since the framework has no impact on the performance, eAM works quite well in comparison to PAM with SB+tree and PAM with UB+tree. However, the large number of insertions and deletions is amortized by the utilization of BB+tree. This is why, the performance of PAM with BB+tree increases significantly.

From the latest experiment we can draw the following conclusions. When the database is large, the framework works for a long time. In this case, the system performance benefits from the optimization techniques used in the framework. When the database is small or heavily modified, the framework works for a short time and does not play a key role in the system. In this case, the system performance depends on the applied index.

\subsubsection{Summary of experiments}
We can summarize all the experiments as follows. We consider three aspects. The first aspect is index choosing for PCM. We applied three different workloads for each index type. The best performance exhibits BB+tree. Thus, we justify the utilization of this index during adaptive merging on PCM.
In the second part of our experiments, we compare the different adaptive merging frameworks. We assume that the database is not modified during adaptive merging. As we can observe, eAM and PAM exhibit similar performance and overpowers the traditional approach (AM).
The last case, where index merging interleaves with database modifications, is more common in the real world.  In this situation, the proposed framework with BB+tree drastically improves the overall system performance.

\section{Related work} \label{rel}

\subsection{Alternative partial indexing strategies} \label{par}
In this paper, we concentrate on the framework for the adaptive merging of the secondary index on PCM. The key parts of the framework are partitions with the data and the adaptive index. Each partition contains secondary index entries sorted by the key. The incoming query copies the entries from the data partitions to the adaptive index.
Worth mentioning is the fact that the indexed data are not removed from the data partitions but are marked as invalid.

Index cracking is an alternative approach (see \cite{SekPattern}, \cite{DBLP:journals/pvldb/HalimIKY12}, \cite{DBLP:journals/vldb/SchuhknechtJD16}, and \cite{DBLP:conf/damon/PirkPIMK14}).
The core idea of index cracking is to consider a query as a hint for data reorganization which can lead to a full index creation. The index is created adaptively, i.e. each range query can potentially make two new sub-partitions using logic very similar to partitioning in quicksort. It leads to a write-heavy workload.
Unfortunately, such a situation is unfavorable for PCM where the write operation is several times more time and energy-consuming than the read operation. As it was proved in \cite{Graefe:2010:SSI:1739041.1739087}, index cracking is not suitable for memory with read-write asymmetry.

In the hybrid cracking (see \cite{IdreosMKG11}, \cite{DBLP:journals/pvldb/HalimIKY12}, \cite{DBLP:conf/dasfaa/XueQZWY13} \cite{DBLP:conf/damon/HaffnerSD18}, and \cite{SekPattern}), the data are stored in the data partitions but, unlike in the case of adaptive merging, they are not initially sorted.  When a range query arrives, the selected data are copied to the sorted final partition where the index is created or modified. After that, the indexed data are deleted from the data partition and the other data in this partition are sorted. The data partitions can be merged if they are too small. The final partition can be also rewritten as a result of sorting or merging. Such an approach leads to write-heavy workload. As a consequence, the overall system performance is decreased on PCM.

Our approach differs from the hybrid cracking in the following aspects. First, we do not rewrite or merge the initial partitions since it would lead to the write overhead on PCM. Instead of removing the already indexed data from the partition, we mark them as invalid using a journal. This technique is significantly faster than a bitmap invalidation (see \cite{RethinkingDatabase}).
Second, we initially sort the data in each partition. In this way, we avoid linear scanning of each partition during the search operation and can apply binary search instead. Moreover, keeping the data sorted is essential to use the journal.
Fortunately, the partitions are initially sorted in the cache which causes a very small overhead.
Third, unlike most approaches, our framework works in the dynamic database environment where updates interleave with range queries.

\subsection{Comparison with other index types} \label{perf}
In our approach, we treat the initial data partitions and the secondary merge index as independent parts of the framework. In this context, we can easily change the underlying index type and utilize the best one for the memory type.

To highlight the optimization, we do the following steps. We start from the traditional adaptive merging with the partitioned B+tree (AM). After that, to make the technique PCM friendly, we consider extended adaptive merging (eAM). This is a traditional adaptive merging that utilizes the B+tree with unsorted leaves (UB+tree) and a bitmap instead of the partitioned B+tree (see \cite{RethinkingDatabase}). At last, to further improve the performance, we use our framework with a new variant of DPTree called BB+tree. This index is equipped with a buffer and two types of sections in the leaves: sorted and unsorted. It turns out that BB+tree is very suitable for adaptive merging where the data are inserted in batches. Additionally, we create a journal that holds the key ranges already stored in the merge index. Such a strategy works drastically better than a bitmap.
Furthermore, we carry out some experiments which compare different index types utilized inside the framework. The experiments clearly show that BB+tree works better than the others. Worth mentioning is the fact that our framework is a very flexible solution. So, the BB+tree can be easily replaced by any other index type in the future.

\section{Conclusion}

The paper faces adaptive merging optimized for PCM.
We implement two methods to overcome PCM limitations in the context of adaptive merging: eAM and PAM. In the first method, we utilize several PCM optimization techniques like the B+tree with unsorted leaves and inner nodes stored in DRAM.
We show that eAM outperforms a traditional approach (AM) by 60\%.
The second method is a framework equipped with the insertion and deletion journal to handle the adaptive merging process on PCM. Although the framework can work with any index type, we propose a new version of DPTree named BB+tree. We observe that PAM is about 20\% faster than eAM in the databases where search queries interleave with data modifications.
For future work, we plan to make further research on BB+tree as a promising candidate for the PCM-optimized index type.



\end{document}